\documentclass[a4paper,11pt]{article}
\pdfoutput=1 

\usepackage{jinstpub} 

\title{\boldmath Module production of the one-arm AFP 3D pixel tracker}

\author[a,b,1]{S.~Grinstein,\note{Corresponding author.}}
\author[a]{E.~Cavallaro,}
\author[a]{M.~Chmeissani,}
\author[c]{O. Dorholt,}
\author[a]{F.~F\"orster,}
\author[a]{J.~Lange,}
\author[a]{I.~Lopez Paz,}
\author[a]{M.~Manna,}
\author[d]{G. Pellegrini,} 
\author[d]{D. Quirion,}
\author[e]{M. Rijssenbeek,}
\author[c]{O. Rohne,} 
\author[f]{and B. Stugu }

\affiliation[a]{Institut de F\'{\i}sica d'Altes Energies (IFAE), \\ 
08193 Bellaterra (Barcelona), Spain}
\affiliation[b]{Instituci\'o Catalana de Recerca i Estudis Avan\c{c}ats (ICREA), \\
Pg. Llu\'{i}s Companys 23, 08010 Barcelona, Spain}
\affiliation[c]{University of Oslo, Department of Physics, \\
Sem S{\ae}landsvei 24 N-0371 Oslo, Norway}
\affiliation[d]{Centro Nacional de Microelectronica (CNM-IMB-CSIC), \\
08193 Bellaterra (Barcelona), Spain }
\affiliation[e]{University of Bergen, Department of Physics and Technology, \\
All{\'e}gaten 55 5007 Bergen, Norway}
\affiliation[f]{State University of New York, Department of Physics and Astronomy,\\
Stony Brook, NY 11794-3800, U.S.A.}

\emailAdd{sgrinstein@ifae.es}

\abstract{The ATLAS Forward Proton (AFP) detector is designed to 
identify events in which one or two protons emerge intact from the LHC 
collisions. AFP will consist of a tracking detector, to measure the 
momentum of the protons, and a time of flight system 
to reduce the background from multiple proton-proton interactions. 
Following an extensive qualification period, 3D silicon pixel sensors 
were selected for the AFP tracker. The sensors were produced at CNM (Barcelona) during 2014. The tracker module assembly 
and quality control was performed at IFAE during 2015. 
The assembly of the first AFP arm and the following installation in 
the LHC tunnel took place in February 2016. 
This paper reviews the fabrication process of the AFP tracker
focusing on the pixel modules.
}


\begin{document}
\maketitle
\flushbottom

\section{Introduction} 
\label{sec:intro}

The ATLAS Forward Proton (AFP)~\cite{afp_tdr} detector extends the physics reach of 
ATLAS~\cite{atlas} by enabling the identification of protons that emerge intact 
and at very low angles from 
the Large Hadron Collider (LHC) proton-proton collisions. 
Such processes are usually 
associated with elastic and diffractive scattering. However, the AFP physics program
ultimately aims to perform searches for new physics by exploring central exclusive
production. 


The AFP detector will consist of high resolution pixel silicon tracker modules 
combined with a time-of-flight (ToF) system, placed at about 210~m from the 
ATLAS interaction point (IP) and at 2-3~mm from the LHC beams.
The tracking detector measures the proton momentum and determines their fractional 
energy loss while the timing system is used to reduce the background from pile-up
events. The detector systems will be placed in Roman Pot 
stations.


%
%

The AFP detector is being installed in two stages. 
In the end-of-year 2015-2016 
LHC shutdown, two tracker units have been placed in two Roman Pot stations
at one side of the IP. 
Since installation, the ``one-arm'' AFP detector has been
operated during special runs with dedicated beam conditions.
The full system with timing detectors, four tracker units
and a total of four Roman Pot stations is 
planned to be completed during the extended end-of-year 2016-2017 shutdown.
To fully exploit the AFP physics potential 
the ultimate goal is to operate the detector  
under regular beam conditions during LHC Run~2, if a safe operation of the 
detector is demonstrated. 

Section~\ref{sec:tracker} describes the one-arm AFP tracker 
and its components.
The sensor productions are presented in section~\ref{sec:sensor_prod}.
The  module construction and qualification is detailed in 
section~\ref{sec:construction}, while the integration and first operation 
of the tracker is briefly reported in section~\ref{sec:integration}.
Summary and conclusions are presented in section~\ref{sec:conclusion}.

\section{Description of the AFP Tracker}
\label{sec:tracker}

The AFP tracker, combined with the magnet system of the LHC 
accelerator, provides the 
momentum measurement of the scattered protons~\cite{afp_tdr}.
In order to fulfill its physics goals, the tracker is required to provide a position
resolution of 10~$\mu$m per four-plane station in the horizontal direction and 30~$\mu$m
in the vertical one~\cite{afp_tdr}. 
Furthermore, the inactive edge of the tracker modules have to be
minimized to increase the detector acceptance. Thus the pixel devices are required to 
have a slim edge of 100-200~$\mu$m.
To be able to operate AFP during normal LHC runs in
Run~2 (expected to deliver 100/fb),
the tracker is required to sustain a non-uniformly distributed fluence 
up to $3 \times 10^{15}$~n$_{\rm{eq}}$/cm$^2$. 
Finally, the AFP tracker has to be able to provide a trigger signal, since in the 
one-arm  operation no timing system is installed. The signal should arrive to 
ATLAS within the level 1 time window. 

Following an extended qualification 
period~\cite{afp_sensor_qualification, afp_tb_paper}, 3D silicon pixel sensors were selected for the AFP tracker
since they were shown to fulfil the above requirements.
The AFP sensors are based on the 3D double-sided technology developed 
by CNM (Barcelona) for the ATLAS IBL project~\cite{IBL, CNM3D}. The 
processed wafers are 100~mm diameter, float zone, p-type, with $<\!100\!>$ crystal
orientation, 230~$\mu$m thickness, and a high resistivity (20~k$\Omega$cm).
Columnar electrodes, 12~$\mu$m wide, are obtained by Deep
Reactive Ion Etching (DRIE) and dopant diffusion from both wafer sides, n$^+$ columns
from the front side, and p$^+$ columns from the back side. Both column types stop about 
30~$\mu$m before reaching the opposite side. The substrate bias is applied from 
the back side. The pixel configuration consists
of two n$^+$ electrodes connected by a metal line to a readout pad, surrounded by six p$^+$ columns 
which are shared with the neighbouring pixels, 
defining the $50\times 250$~$\mu$m$^2$ pixel geometry. 
The n$^+$ columns are isolated 
with p-stop implants. The edge isolation is accomplished with a combination of a n$^+$ 3D guard 
ring, which is grounded, and p$^+$ fences, which are at the high voltage (HV) potential from the 
ohmic side. The AFP sensors present a 180~$\mu$m slim inactive 
edge on the side that faces the LHC beam.

The front-end electronics is the FE-I4B chip~\cite{fei4b, fei4a} developed by the ATLAS 
collaboration for the IBL detector. The sensors are DC coupled to the chip with negative 
charge collection. Each readout channel contains an independent amplification stage with 
adjustable shaping, followed by a discriminator with independently adjustable threshold. 
The chip operates 
with a 40 MHz externally supplied clock. The time over threshold (ToT) with 4-bit resolution 
together with the firing time are stored for a latency interval until a trigger decision is 
taken. In the case of any discriminator being above threshold, the FE-I4B chip issues
a fast signal through a dedicated HitOr pad. These signals are used to generate a trigger signal 
from AFP to ATLAS.

\begin{figure}[htbp]
\begin{center}
\includegraphics[width=.4\textwidth]{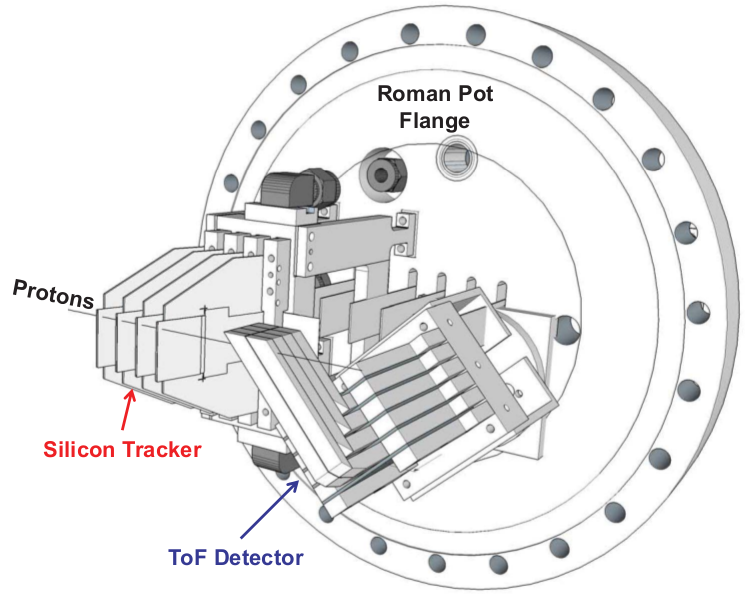}
\qquad
\includegraphics[width=.3\textwidth]{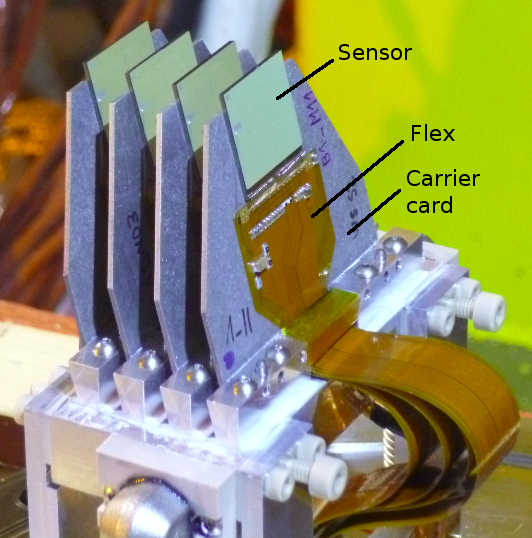}
\caption{\label{fig:tracker} Design of the AFP detector including tracking and time 
of flight systems (left). The diffractive protons arrive from the left. The tracker 
modules mounted in the support frame (right). In this picture the slim edge side is 
facing up.}
\end{center}
\end{figure}

As already mentioned, the one-arm AFP system consists of two tracking stations placed in 
Roman Pots at 205~m and 217~m from the IP, 
that approach the beam horizontally. Each station is composed of four pixel
modules arranged such that the short pixel direction 
of the sensor is horizontal to the LHC floor.
Each module includes a 3D sensor interconnected to a single FE-I4B front-end chip,
glued to a support holder and wire-bonded to a flexible printed circuit (referred to as flex).
The modules are placed with a pitch of 9~mm and tilted by 14$^\circ$ with respect to the 
beam direction, to enhance hit reconstruction 
efficiency and horizontal position resolution 
(see figure~\ref{fig:tracker}).

\section{Sensor Productions}
\label{sec:sensor_prod}

The first CNM production of 3D sensors for AFP (run 6682), which started with 13 wafers,
finished in July 2014. 
The fabrication mask set was identical to the IBL one.
Each wafer included 8 3D sensors compatible with the FE-I4B chip.
However, mainly due to damage at the wafer edge introduced during the DRIE process,
8 wafers broke during production. 
The remaining wafers were processed for under bump metalization at IZM (Germany) and diced
at CNM with a standard diamond saw to 180~$\mu$m on the 
side facing the LHC beam.

The electrical quality of the sensors was determined through the measurement
of the current-voltage (I-V) curves,
which were obtained for each diced sensor using a probe station at room temperature. 
The n-side of the sensor was placed in contact
with the grounded chuck via the under bump metallization, while the p-side was 
connected to the bias potential.
This procedure does not ensure that all pixels are correctly biased (due to
irregularities in the bumps, dust deposited in the chuck, sensor bow, etc), and a better method, explained below, was implemented in a successive sensor production.
Given that the IBL 3D sensors showed excellent hit reconstruction 
efficiency even at low voltages~\cite{afp_tb_paper}, 
the sensors were classified as green (best quality) and yellow (medium quality)
if the leakage current was below 10~$\mu$A at 20~V and 10~V respectively. 
The rest of the sensors were classified as red. 
%
The first production resulted in only 9 (5) green (yellow) sensors. 
Follow-up studies performed with a 
scanning electron microscope indicated that during the etching process 
damage was introduced to the column sidewall, which 
could partially explain the poor sensor yield.
The low number of green sensors implied that 
problems that might result in the loss of sensors during the module assembly 
process (from bump-bonding to module
testing) had to be minimized in order to produce enough tracking devices
for the one-arm installation.

%
 
In order to test if the poor electrical performance of the sensors could 
also be related to an excessive p-stop implantation, 
several sensors with low breakdown voltage were irradiated 
at the TRIGA reactor at JSI Ljubljana with neutrons to fluences 
of 1, 10 and $100 \times 10^{12}\,\rm{n}_{\rm{eq}}$/cm$^2$ and annealed for a week at
room temperature. The intention was to investigate if acceptor removal
reduced the electric field in the area of the p-stops. 
Three of the 7 red sensors that were irradiated to $1 \times 10^{14}\,\rm{n}_{\rm{eq}}$/cm$^2$  
showed a clear improvement in the electrical behaviour, with
a break down voltage (V$_{\textrm{bd}}$) that increased from a few volts to 
20-35~V. One of these sensors was eventually mounted and installed in AFP (see next section).

\begin{figure}[htbp]
\begin{center}
\includegraphics[width=.5\textwidth]{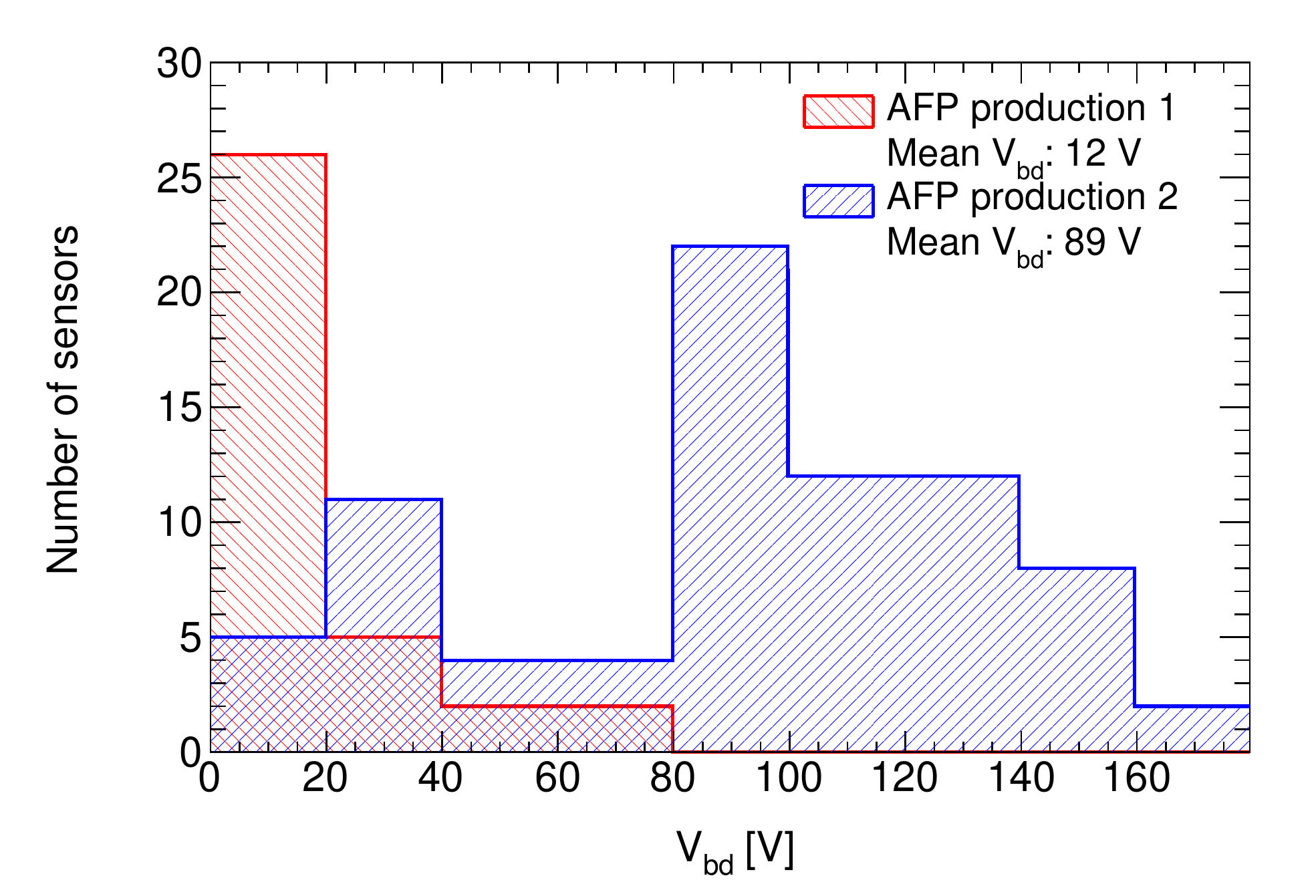}
\caption{\label{fig:3d_production_yield} Break down voltage of 3D sensors from the 
first and second AFP productions.}
\end{center}
\end{figure}

Due to the low yield of the first AFP sensor production, CNM launched a second run
(7945) after improving the fabrication process. The edges of the wafers
were protected during the DRIE 
and the process parameters were optimized to reduce damage to the column sidewall. 
To refine the sensor selection process, the second production included 
a temporary metal layer that connected all the pixels. 
In combination with the pad of 
the 3D guard ring, the metal layer allowed an accurate determination of the I-V curve.
The layer was removed before dicing by chemical etching.
The second 
production finished in March 2016, with a distinctly improved yield: only 2 of the 12 wafers 
were lost during fabrication, while 68 of the 80 remaining sensors were of green quality.  
Table~\ref{tab:yields} summarizes the yields
of the two AFP production runs while figure~\ref{fig:3d_production_yield}
compares the break down voltages.

\begin{table}
\begin{center}
\caption{\label{tab:yields} Yield of the first and second sensor productions for AFP. After 
the fabrication process was optimized, a substantial improvement in both wafer and
sensor yields was achieved.}
\smallskip
\begin{tabular}{|c|c|c|c|c|}
\hline
Production & Wafer & Good   & Sensor & Good \\
run        & yield & wafers & yield  & sensors \\
\hline
AFP 1 (6682) & 38\% & 5 & 23\%  & 9 \\
AFP 2 (7945) & 83\% & 10 & 85\% & 68 \\
\hline
\end{tabular}
\end{center}
\end{table}

\section{Module construction and quality assurance}
\label{sec:construction}


The module production for the one-arm AFP detector needed to achieve an excellent yield, 
given the limited
number of good sensors available from the first sensor production.
In order to avoid mistakes that could affect several modules, the assembly process was
serialized, with at most two modules being mounted at the same time. The fact that 
all the assembly process steps, from hybridization 
(connection between sensor and readout electronics)
to quality assurance (QA), were done
in-house at IFAE, ensured that even this approach was relatively fast. The typical
assembly time was $2$--$3$ days per module. 

To gain experience with the
assembly process, low quality sensors were selected for the first modules, while the best ones were processed last. 
Bump-bonding was performed by IFAE with a S\"uss Microtech FC-150 flip-chip 
bonder machine with a reflow arm.
During the bonding cycle the FE-I4B chips (with SnAg solder bumps) and the 
sensors were aligned, heated to 260~$^{\circ}$C for a
short period (about a minute) and pressed together lightly. 
In order to reduce the number of unconnected bumps due to bowing
during the hybridization process, the thickness of the chips was kept 
at 700~$\mu$m (original wafer thickness).
After flip-chip, the assemblies were processed in a fluxless formic acid
reflow oven.

The bare assemblies were inspected with a high resolution (sub-micron) 
X-ray machine to discard devices with a large number of disconnected bumps. 
During the full production, 24 sensors of run 6682 were bump-bonded, though
many of these devices were used to improve the assembly process. Only
one device presented an area of disconnected pixels. The shape of this area 
indicated that 
the FE-I4B bumps were probably damaged during handling prior to
the hybridization cycle. 
The I-V curves of the bare assemblies were measured with a probe station
before the modules were fully mounted. Good correlation with the 
sensor pre-assembly measurements from the probe station was found.

The bare assemblies were glued to aluminium-carbon fiber carrier
cards, designed by SUNY at Stony~Brook and produced at Novapack. 
A combination of double sided Tesa tape and 
Araldite 2011 glue was used, the
later being a radiation hard glue, already used in ATLAS for the
IBL detector. The
placement of the bare assemblies on the carrier cards was done
with a pick-and-place machine 
using two alignment marks located in the cards.
However, the distance between the assembly edge and the alignment marks 
in the long pixel direction was 2300~$\mu$m, which prevented an optimal
alignment of the assembly. Furthermore, the 1~mm thick cards were found to 
present a $\approx$100~$\mu$m bow caused by the thinning to 0.5~mm
of the region of the card where the FE-I4B chip is placed.
This bowing hampered the correct placing of the modules during the
placing step. The effect was reduced by increasing the thickness 
of the layer of tape. 
The final modules achieved an average placement
precision of about 10~(30)~$\mu$m in the short (long) pixel direction.

The flex circuit, designed by the University of Oslo, was glued to the 
card and the module production
completed by wire-bonding. Long wires ($\approx 2$~cm) connected the
HV pad of the flex to the sensor passivation opening on 
the p-side.
The wirebonding quality was checked with wire-pull tests which
indicated that 8-11~g were needed to break the bonds. The
bonds broke at the heel during these tests.

The assembled modules were tested with the USBpix~\cite{usbpix} 
and the HSIO-II~\cite{hsio2} readout systems.
The devices were tuned to  to a threshold of 2~ke$^-$ 
and a ToT of 10 (in units of 25~ns clock cycles) at an injected charge 
of 20~ke$^-$, which corresponds to the values later used during the detector 
operation periods in 2016.  The calibration was performed through 
the internal charge-injection mechanism of the FE-I4B chip. The calibration parameters 
that control the injection 
(injection capacitance, slope of the transfer function of the injection voltage DAC
and reference current setting)
of each chip was previously determined at wafer level. 
The noise level of each assembly was extracted during the threshold calibration procedure.
After calibration, tests with an external source were performed to verify charge collection and 
the connectivity of the front-end chip to the sensor. As already mentioned,
only one device showed a significant number of disconnected bumps (>$1\%$). 

In order to check the mechanical stability of the assembly process, a module
was subjected to an environmental stress test of multiple 
one hour thermal cycles from 
$-20\,^\circ$C to $40\,^\circ$C during one full day. The module was not 
operated during the cycles.
Afterwards, the threshold and ToT calibrations, and the noise level, were
compared with the initial results.  No variations were found. Furthermore,
no mechanical degradation was identified during an optical inspection.
At a later stage, further thermal cycle tests
were performed with four modules assembled for the second AFP arm. No electrical 
nor mechanical degradation was observed. 

In total 17 modules, presented in table~\ref{table:afp_modules}, 
were fully assembled.
Two devices did not pass the QA procedure,
since the communication with the front-end could not be established. Most of the other modules
showed good performance with noise levels compatible with IBL modules. 
Module M01, with a sensor of very poor quality, and 
module M17 which presented a floating guard ring (the bumps that connect the 
guard ring to ground were removed before bump-bonding), showed a significantly
higher noise.
The overall assembly yield (including bump-bonding) was of 82~$\%$ (14/17).

 
%

\begin{table}
\begin{center}
\caption{\label{table:afp_modules} 
Summary of the AFP tracker module production. The sensor quality before and after irradiation is shown for pre-irradiated sensors. For two modules
the communication with the front-end could not be established. Module 
M10 was operated at 0~V due to a shorted HV line. Modules M01 to M11 were
mounted at the time of one-arm tracker assembly at CERN.}
\smallskip
\small
\begin{tabular}{|c|c|c|c|c|c|c|}
\hline
Module & Pre-irrad. & Sensor & Disconnected & Noise & Installed & Operational \\
       &    ($10^{12}\,\rm{n}_{\rm{eq}}$/cm$^2$)        
                    & quality& bumps        & (e$^-$)  &        & Voltage (V)\\
\hline
M01    & No         & red    & No & 320$\pm$37    & No  & --- \\
M02 &    No         & red    & --- & No comm.           & No  & --- \\
M03 &    No         & yellow & No  & 202$\pm$18    & Far  & 10 \\
M04 &    100        & red/green & No  & 163$\pm$11 & Far  & 30 \\
M05 &    No         & yellow & No  & 162$\pm$11 & Near  & 5 \\
M06 &    No         & yellow & No  & 160$\pm$10 & Near  & 10 \\
M07 &    No         & green  & 15\%& 154$\pm$13 & No  & --- \\
M08 &    No         & green  & No  & 160$\pm$11 & Near  & 10 \\
M09 &    No         & yellow & No  & No comm.   & No  & --- \\
M10 &    No         & green  & No  & 156$\pm$11 & Far  & 0 \\
M11 &    No         & green  & No  & 166$\pm$11 & Far   & 5 \\
M12 &    No         & green  & No  & 169$\pm$11 & No   & --- \\
M13 &    No         & green  & No  & 171$\pm$11 & No   & --- \\
M14 &    No         & green  & No  & 165$\pm$11 & No   & --- \\
M15 &    No         & green  & No  & 175$\pm$11 & No   & --- \\
M16 &    10         & red/green  & No  & 175$\pm$12 & No   & --- \\
M17 &    No         & green  & No  & 262$\pm$20 & No   & --- \\
\hline

\hline
\end{tabular}
\end{center}
\end{table}


\section{AFP tracker integration and operation}
\label{sec:integration}

At the time of the one-arm AFP detector assembly at CERN, 4 tracker modules
with green sensors were available, but since one had disconnected bumps,
3 of them were selected for installation. Another 3 yellow modules
and a sensor which was pre-irradiated were also selected. 
The seven modules were shipped to CERN and fully re-tested 
prior to installation in the LHC tunnel in February 2016.
During the initial testing of the system, it became apparent that
the HV line of one module was shorted. This module was operated
at 0~V. The one-arm tracker consists of three modules in the near station 
and four, but one without HV, in the far station. 
The detector temperature varies between $-5\,^\circ$C and $0\,^\circ$C
depending on operation conditions. 

The LHC accelerator ramp-up period (April-May 2016) allowed
to integrate AFP with the ATLAS DAQ system. During the
one bunch crossing LHC operation mode, AFP was timed-in with respect to ATLAS,
ensuring that the AFP readout was within the ATLAS latency. 
LHC configurations with low number of bunch crossings
were also used to study the positioning procedure of the Roman Pot
system.

The AFP tracker was fully commissioned~\cite{ivan_elba} 
in the first half of 2016 and 
diffractive events collected during dedicated low luminosity runs 
in August and October. 
Multiple high luminosity runs were also carried out through the year.
Valuable experience in detector operation, performance and the interaction
of AFP with other ATLAS sub-detector systems was collected during 
these runs.

\section{Conclusions and outlook}
\label{sec:conclusion}

AFP aims to extend the physics reach of ATLAS by identifying
events with one or two intact protons scattered in the forward 
direction. The one-arm
AFP detector includes a tracking system with 3D pixel sensors.
The first sensor production at CNM suffered from low yield.
However, due to the excellent turnout of the module assembly
process, seven tracker planes were installed in two Roman Pot
stations on one side of the ATLAS IP in the LHC shutdown of 2015-2016.
The detector successfully collected LHC proton-proton collision data 
during dedicated runs in 2016. 

After the sensor production process was optimized, the recently
completed second sensor production showed a significantly improved yield.
New tracker modules are being assembled for the full two-arm AFP
detector that will also include the ToF system and which is
foreseen to be installed in the shutdown of 2016-2017.

\acknowledgments

This work was partially funded by: the MINECO, Spanish Government, under grants 
FPA2015-69260-C3-2-R, FPA2015-69260-C3-3-R
and SEV-2012-0234 (Severo Ochoa excellence programme); and the European 
Union's Horizon 2020 Research and Innovation programme under Grant Agreement no. 654168.




\begin{thebibliography}{99}

\bibitem{afp_tdr}
{ATLAS} Collaboration, {\em {Technical Design Report for the
  ATLAS Forward Proton Detector}}, Tech. Rep. CERN-LHCC-2015-009.
  ATLAS-TDR-024, CERN, Geneva, May, 2015.

\bibitem{atlas}
{ATLAS} Collaboration, {\em {The ATLAS Experiment at the CERN Large Hadron
  Collider}},
{JINST {\bfseries 3} (2008) S08003}.

\bibitem{afp_sensor_qualification}
J.~Lange, E.~Cavallaro, S.~Grinstein, and I.~Lopez~Paz, {\em 3D silicon pixel
  detectors for the ATLAS Forward Physics experiment},
  {JINST {\bfseries 10}
  (2015) C03031}.

\bibitem{afp_tb_paper}
J.~Lange, et al., \emph{Beam tests of an integrated prototype of the ATLAS Forward Proton detector}, \emph{JINST}, {\bf 11} (2016) P09005.

\bibitem{IBL}
ATLAS Collaboration, {\em {ATLAS Insertable B-Layer Technical Design Report}}, 
Tech. Rep. CERN-LHCC-2010-013. ATLAS-TDR-19, CERN, Geneva, September, 2010.


\bibitem{CNM3D}
G.~Pellegrini {et~al.}, {\em 3D double sided detector fabrication at IMB-CNM},
{Nucl. Instrum. Meth.  {\bfseries {A 699}} (2013) 27}.

\bibitem{fei4b} 
V. Zivkovic, et al., \emph{The FE-I4 pixel readout system-on-chip resubmission 
for the insertable B-Layer project}, \emph{JINST}, {\bf 7} (2012), C02050.

\bibitem{fei4a}
M.~Garcia-Sciveres {et~al.}, {\em The FE-I4 pixel readout integrated circuit},
  \href{http://dx.doi.org/10.1016/j.nima.2010.04.101}{Nucl. Instrum. Meth.
  {\bfseries A 636} (2011) S155}.

  
\bibitem{usbpix} M.~Backhaus, et al., {\em Development of a versatile and modular test 
system for ATLAS hybrid pixel detectors}, {Nucl. Instrum. Meth.
  {\bfseries {A 650}} (2011) 37}.

\bibitem{hsio2} 
M.~Wittgen {et~al.}, {\em A Reconfigurable Cluster Element (RCE) DAQ Test Stand
  for the ATLAS Pixel Detector Upgrade}, 
\newblock
 \href{https://indico.cern.ch/event/83060/session/16/contribution/126/material/slides/0.pdf}
  {Topical Workshop on Electronics for Particle Physics (TWEPP), Aachen, Germany, September 2010}.

\bibitem{ivan_elba} 
I.~Lopez~Paz, {\em The AFP Silicon Tracker},
\newblock
 \href{http://indico.cern.ch/event/505807/}
{3rd Elba Workshop on Forward Physics @ LHC Energy, Elba, Italy, May 2016.}







\end{thebibliography}
\end{document}